# Sparse Graphs for Belief Propagation Decoding of Polar Codes

Sebastian Cammerer, Moustafa Ebada, Ahmed Elkelesh and Stephan ten Brink
Institute of Telecommunications, Pfaffenwaldring 47, University of Stuttgart, 70569 Stuttgart, Germany
{cammerer,ebada,elkelesh,tenbrink}@inue.uni-stuttgart.de

*Abstract*—We describe a novel approach to interpret a polar code as a low-density parity-check (LDPC)-like code with an underlying *sparse* decoding graph. This sparse graph is based on the encoding factor graph of polar codes and is suitable for *conventional* belief propagation (BP) decoding. We discuss several pruning techniques based on the check node decoder (CND) and variable node decoder (VND) update equations, significantly reducing the size (i.e., decoding complexity) of the parity-check matrix. As a result, iterative polar decoding can then be conducted on a sparse graph, akin to the traditional well-established LDPC decoding, e.g., using a *fully parallel* sum-product algorithm (SPA). This facilitates the systematic analysis and design of polar codes using the well-established tools known from analyzing LDPC codes. We show that the proposed iterative polar decoder has a negligible performance loss for short-to-intermediate codelengths compared to Arıkan's original BP decoder. Finally, the proposed decoder is shown to benefit from both reduced complexity and reduced memory requirements and, thus, is more suitable for hardware implementations.

## I. Introduction

Polar codes introduced in [1] are the first type of channel codes that are theoretically proven to achieve the channel capacity for infinite length codes under successive cancellation (SC) decoding. Unfortunately, finite length polar codes suffer from poor error-rate performance under SC decoding. A belief propagation (BP) decoding algorithm based on [2] was proposed in [3] to enhance the finite length performance. Later, a successive cancellation list (SCL) decoder was proposed achieving the maximum likelihood (ML) bound [4].

Polar codes have been recently adopted to the 5G standard as a part of the uplink control channel, thus, practical decoding algorithms of polar codes have become a very attractive topic. One major drawback of SC and SCL decoding is a long decoding latency due to its serial decoding nature, in which the information bits are decoded one at a time, reducing the overall throughput of the decoder. Besides, the SCL decoder does not provide soft-in/soft-out information and thus not suitable for iterative decoding and detection. On the other hand, the BP decoder does not suffer from the above mentioned issues, thus it is a good candidate for high data rate demanding applications (i.e., from an implementation perspective).

The first BP decoder proposed for polar codes is performed over a factor graph corresponding to the *generator matrix* of the code [3]. A straightforward conversion from the generator matrix-based factor graph to a parity-check matrix-based factor

This work has been supported by DFG, Germany, under grant BR 3205/5-1.

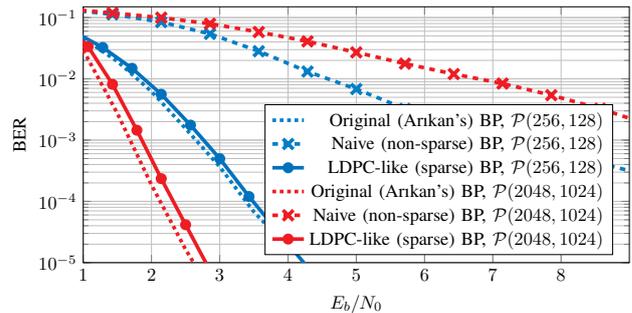

Fig. 1: BER comparison between Arıkan's original BP decoder, BP decoder performed over the naive $\mathbf{H}$-based dense factor graph, and our proposed LDPC-like BP decoder performed over the pruned factor graph for $\mathcal{P}(256, 128)$ and $\mathcal{P}(2048, 1024)$ codes; $N_{it,max} = 200$.

graph [5] leads to a *dense* (i.e., non-sparse) parity-check matrix with many short cycles as shown in Fig. 2. Thus, the traditional BP algorithm as used in low-density parity-check (LDPC) decoding, with variable node decoder (VND) and check node decoder (CND), fails over the dense graph corresponding to the resulting dense parity-check matrix (see Fig. 1). To allow iterative decoding over a suitable parity-check matrix, corresponding graph pruning steps have been reported in [6]. In this work, we propose a novel approach for finding a sparse parity-check matrix[1] of a corresponding LDPC code suitable for iterative decoding after pruning the redundant variable nodes (VN) and check nodes (CN). Thus, a polar code can be decoded with a slightly modified generic BP decoder (e.g., an LDPC BP decoder) with negligible bit error rate (BER) performance loss (refer to Fig. 5). By viewing polar codes as an LDPC-like code, the LDPC design methods or analysis tools (e.g., EXIT charts and density evolution) may become applicable in the context of polar codes. This will facilitate the hard task of systematically analyzing polar codes under BP decoding and may be useful in finding the best polar code construction (i.e., the frozen bits) algorithm for polar codes tailored to iterative decoding. This opens door to new and, probably, more efficient hardware implementation, given the latency reduction provided here and given that polar codes can be, surprisingly, decoded iteratively on a sparse $\mathbf{H}$-based factor graph of reduced size using the conventional well-established

---

[1] Strictly speaking, the proposed matrix $\mathbf{H}$ is not a parity-check matrix of the polar code, as $\mathbf{H}\tilde{\mathbf{x}}$ requires an extended codeword $\tilde{\mathbf{x}}$. However, for readability and with abuse of some notation we denote $\mathbf{H}$ as the parity-check matrix of a corresponding LDPC-like code.

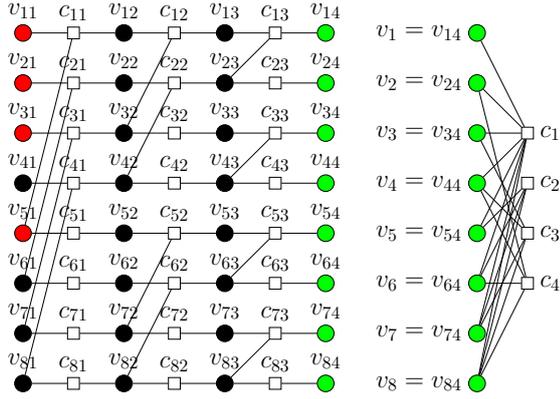

Fig. 2: (*Left*) Original (Arıkan's) BP decoder factor graph for a $\mathcal{P}(8,4)$-code with $\mathbb{A} = \{4,6,7,8\}$. (*Right*) Corresponding dense factor graph, i.e., non-sparse $\mathbf{H}$, based on [5, Lemma 1]. LDPC decoders.

## II. CONVENTIONAL BP DECODING OF POLAR CODES

Polar codes are based on the concept of channel polarization in which a set of $N$ identical channels are combined based on a $2 \times 2$ kernel, and the output of $N$ synthesized bit channels shows a polarization behavior. Then, the best $k$ synthesized bit channels are used for data transmission and the remaining $N - k$ bit channels are set to a frozen known value (w.l.o.g. we assume 0 throughout this work) to both the transmitter and receiver. For the rest of this work, such a polar code is denoted by $\mathcal{P}(N,k)$. The information set and the frozen set are denoted by $\mathbb{A}$ and $\bar{\mathbb{A}}$, respectively. Throughout this work, the polar code is constructed based on Arıkan's Bhattacharyya bounds of bit channels [1] with a design SNR of 0.6 dB. The encoding is based on the generator matrix $\mathbf{G}$, i.e., a polar codeword is $\mathbf{x} = \mathbf{u} \cdot \mathbf{G}$, where $\mathbf{u}_\mathbb{A}$ contains the useful information bits and the remaining positions of $\mathbf{u}$ are frozen. Besides, systematic encoding is also possible with affordable complexity, offering a significant BER improvement [7]. For the rest of this work, systematic polar encoding is used.

Arıkan's original BP decoder in [3] is based on Forney's BP decoder for Reed–Muller (RM) codes using factor graph representations of the generator matrix $\mathbf{G}$ [2], where subsequent iterations are conducted on the polar code *encoding* factor graph [3]. The original polar BP decoder is briefly revisited:

*1) Factor graph:* The original polar BP decoder factor graph of a $\mathcal{P}(8,4)$-code, Fig. 2, consists of $n = \log_2 N$ stages, where each stage contains two types of nodes. The factor graph contains a set of $N \cdot (n+1)$ VNs $\{v_{ij}\}$ and a set of $N \cdot n$ CNs $\{c_{ik}\}$, where $1 \leq i \leq N$, $1 \leq j \leq n+1$, $1 \leq k \leq n$.

*2) Iterative message passing:* For the sake of simplicity, we use the terminology as familiar from LDPC decoding and follow the basic definitions from [8]. We refer the interested reader to [3] for more details. Let $x_{v \to c}$ denote the log-likelihood ratios (LLR) message from VN $v$ to CN $c$ and $y_{c \to v}$ denote the LLR message from CN $c$ to VN $v$, respectively. Further, we define $\mathcal{V}_c$ as the set of connected VNs to CN $c$ and $\mathcal{C}_v$ as the set of connected CNs to VN $v$. The CN update is computed as

$$y_{c \to v} = \prod_{v' \in \mathcal{V}_c / v} \text{sign}\left(x_{v' \to c}\right) \cdot \phi \left( \sum_{v' \in \mathcal{V}_c / v} \phi \left(|x_{v' \to c}|\right) \right) \quad (1)$$

where $\phi(x) = \phi^{-1}(x) = -\log\left(\tanh\left(\frac{x}{2}\right)\right)$. For the following part, it is instructive to realize that $\phi(0) \to \infty$ and $\phi(\infty) \to 0$. The VN update is

$$x_{v \to c} = L_v^{init} + \sum_{c' \in \{\mathcal{C}_v/c\}} y_{c' \to v}. \quad (2)$$

where $L_v^{init}$ denotes the a priori knowledge of the VN $v$.

The VNs at the leftmost stage $\{v_{i,1}\}$, $1 \leq i \leq N$ are initialized (i.e., $L_{v_{i,1}}^{init}$ is set accordingly) according to the a priori knowledge available to the decoder (i.e., the frozen bits), whereas the VNs at the rightmost stage $\{v_{i,n+1}\}$, $1 \leq i \leq N$ are initialized according to the channel output (i.e., the received LLRs $L_{ch}$). All remaining VNs initially have no a priori knowledge, thus $L_{init} = 0$.

*3) Scheduling, termination and stopping conditions:* Instead of a fully parallel message update, the nodes are updated stage-by-stage. A single full decoding iteration, thus, describes the consecutive activation of stage 1 to $n$ (forward) and from $n$ to 1 (backward).

Traditionally, BP decoding is terminated when reaching a pre-defined maximum number of BP iterations $N_{it,max}$ or achieving a specific pre-defined stopping criterion, e.g., all CNs are satisfied (see [9] for details on stopping conditions of polar BP decoding).

The parity-check matrix $\mathbf{H}$ of a polar code with generator matrix $\mathbf{G}$ can be constructed from the columns of $\mathbf{G}$ with indices in $\bar{\mathbb{A}}$, where $\bar{\mathbb{A}}$ denotes the set of frozen indices [5, Lemma 1]. The resulting $\mathbf{H}$-matrix, shown in Fig. 2, is a highly dense parity-check matrix (see Table I) with a maximum check node degree of $N$. Consequently, this leads to a decoding failure if traditional decoding algorithms (e.g., sum-product algorithm (SPA) decoding) are performed on this dense $\mathbf{H}$, as shown in Fig. 1. The parity-check matrix of the sparse factor graph was introduced in [5], later termed *adjacency matrix* in [10], and was used to construct the Linear Programming (LP) polytope which defines the LP decoder proposed in [5].

Table I: Density of non-sparse $\mathbf{H}$-matrix of polar codes based on [5].

| Code | $\mathcal{P}(256,128)$ | $\mathcal{P}(2048,1024)$ | $\mathcal{P}(8192,4096)$ |
|---|---|---|---|
| den($\mathbf{H}$) | 16.31% | 7.11% | 4.06% |

## III. PRUNING TECHNIQUES FOR POLAR FACTOR GRAPHS

As shown in Fig. 2, Arıkan's BP decoding graph can be re-drawn into a bipartite graph consisting of VNs $\mathcal{V}$ and CNs $\mathcal{C}$ [5]. By neglecting the stage-wise decoding scheduling, the graph represents an *LDPC-like* code structure, or a particularly constrained LDPC code, where the last $N$ bits of the code represent the polar codeword. Besides, there is no channel input over the first $|\mathcal{V}| - N$ bits. However, the $\mathbf{H}$-matrix corresponding to the LDPC-like code is quite large (i.e., matrix size is $N \cdot \log_2 N \times N \cdot (\log_2 N + 1)$ compared to $N - k \times N$ for the dense $\mathbf{H}$ of the original polar code, cf. Table II).

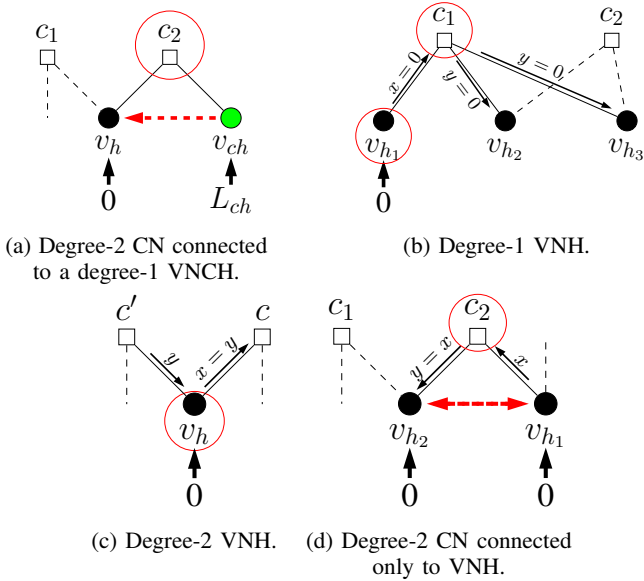

Fig. 3: Special cases of polar factor graph nodes.

We now show that, only based on Eq. (1) and (2), effective pruning of this graph can be applied, leading to a reduced LDPC-like code, that can be decoded efficiently with a conventional LDPC BP decoder. Similar ideas exist for LP decoding of polar codes [10]. We define different kinds of VNs, that only differ in the effective underlying channel $L_{ch}$:

1) Channel variable nodes (VNCH) $v_{ch} \in \mathcal{V}_{ch}$ which represent a codeword position, i.e., they are initialized with the corresponding $L_{ch}$ (green VNs in Figures 2-4).
2) Hidden variable nodes (VNH) $v_h \in \mathcal{V}_h$ which define the intermediate nodes without any prior knowledge, i.e., $L_{ch} = 0$ (black VNs in Figures 2-4).
3) Frozen VN $v_f \in \mathcal{V}_f$ at frozen positions are assumed to see $L_{ch} = \infty$ (red VNs in Figures 2-4).

As we assume systematic encoding[2], the estimated information bits $\hat{\mathbf{u}}$ can be directly extracted from the VNCH.

We identify several common scenarios which occur during the decoding. We want to emphasize that these cases are very specific for the underlying polar code structure and can not be found in LDPC codes (i.e., there are no VNH). However, these observations facilitate finding a compact graph representation.

*1) Frozen nodes:* According to Eq. (2), a frozen VN $v_f$ always propagates $x_{v_f \to c} = \infty$, independently of its received messages. Thus, the corresponding CN update is

$$y_{c \to v} = \underbrace{\text{sign}(x_{v_f \to c})}_{=1} \cdot \prod_{v' \in \mathcal{V}_c / \{v, v_f\}} \text{sign}(x_{v' \to c})$$
$$\cdot \phi \left( \underbrace{\phi(|x_{v_f \to c}|)}_{\to 0} + \sum_{v' \in \mathcal{V}_c / \{v, v_f\}} \cdot \phi(|x_{v' \to c}|) \right)$$

[2]Non-systematic encoding can be applied straightforwardly by an additional re-encoding step after the decoding.

and, as a result, the frozen node can simply be removed from the graph as it never contributes to the decoding process. Based on this observation, the size of the $\mathbf{H}$-matrix may vary depending on the code construction (i.e., $\mathbb{A}$).

*2) Degree-1 CN:* The underlying parity-check equation of a degree-1 CN can only be fulfilled if the connected VN represents the bit "0". From Eq. (1)

$$y_{c \to v} = \phi(0) \to \infty.$$

Thus, the degree-1 CNs and the connected VN can be removed from the graph (cf. frozen bits).

*3) Degree-1 VNCH and degree-2 CN:* Obviously, a VNCH of degree-1 simply propagates the received $L_{ch}$ into the connected CN and never changes its message $x$. In case the other connected VN is a VNH, the incoming message at this node is always equal to $L_{ch}$ from the VNCH. Therefore, the CN can be removed and the VNH must be replaced by the corresponding VNCH. This is depicted in Fig. 3a.

*4) Degree-1 VNH:* A VNH of degree-1 ($v_{h_i}$) always returns $x_{v_{h_i} \to c} = 0$ to the CN (see Fig. 3b). Using Eq. (1) we find

$$y_{c \to v} = \underbrace{\text{sign}(x_{v_{h_i} \to c})}_{=1} \cdot \prod_{v' \in \mathcal{V}_c / \{v, v_{h_i}\}} \text{sign}(x_{v' \to c})$$
$$\cdot \phi \left( \underbrace{\phi(|x_{v_{h_i} \to c}|)}_{\to \infty} + \sum_{v' \in \mathcal{V}_c / \{v, v_{h_i}\}} \phi(|x_{v' \to c}|) \right) \to 0,$$

and thus, the CN never contributes to the decoding and can be removed from the graph together with the degree-1 VNH.

*5) Degree-2 VNH:* From Eq. (2) it follows that VNH of degree-2 are simple *feed-forward* nodes (cf. Fig. 3c) and the corresponding VNH update equation simplifies to

$$x_{v_h \to c} = \underbrace{L_{v_h}^{init}}_{=0} + y_{c' \to v_h}.$$

Thus, the VNH can be removed by merging the two CNs that are connected to it[3].

*6) Degree-2 CN:* A CN of degree-2, which is only connected to VNH (see Fig. 3d) defines a *forwarding* node and can thus be removed by merging the two VNHs into one VNH.

Based on all that, we define a pruning algorithm (see Algorithm 1), similar to [6], which simplifies the decoding graph without a significant impact on the decoding performance. The algorithm simplifies the graph iteratively until the size of $\mathbf{H}$ does not change anymore. It is worth to keep in mind that further pruning opportunities may be possible as we here define the baseline schemes. Fig. 4 illustrates the pruning of a $\mathcal{P}(8,4)$-code. The $\mathbf{H}$-matrix can be pruned from $24 \times 32$ to $5 \times 9$. We provide the source code online to facilitate reproducing the results for further investigation and more pruning ideas [11]. Following this algorithm, the dimensions (and density) of the sparse **H**-matrix and the degree profiles before pruning are compared to that after pruning for different code lengths, as given in Table II. It can be observed that the pruning algorithm leads to a significant matrix size, or graph size, reduction (with a reduction factor of $\approx 83-85\%$ after pruning), while maintaining the sparsity of the **H**-matrix.

[3]Remark: this slightly changes the scheduling, as the messages are delayed by one iteration through the VNH. However, we did not observe any significant impact due to this effect.

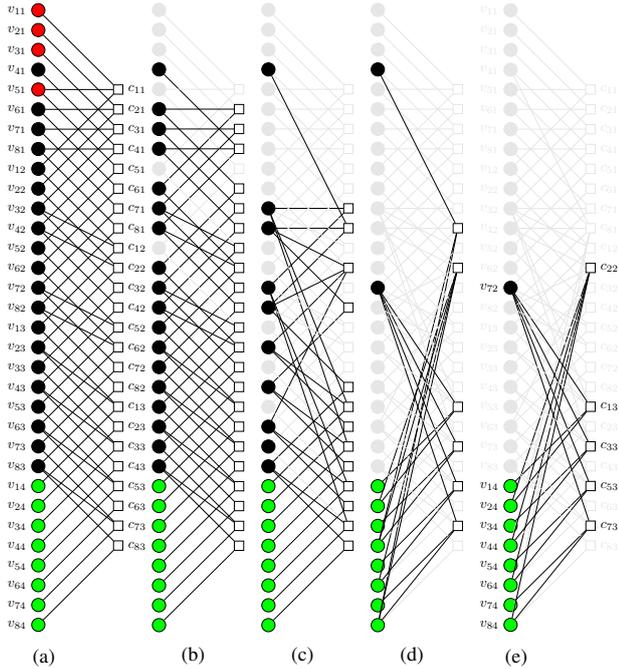

Fig. 4: Pruning of the $\mathcal{P}(8,4)$-code graph with $\mathbb{A} = \{4,6,7,8\}$.

One might expect that the pruning process reduces the portion of the degree-1 VNs due to removing the degree-1 VNHs. However, while pruning the graph, the degree-1 VNCHs are not all removed/altered and thus become more dominant in the degree profile of the reduced **H**-matrix, as can be seen from the variable and check degree distribution from an edge-perspective of the **H**-matrix ($\lambda(Z)$ and $\rho(Z)$) as provided in Table II. It is also observed that the portions of degree-2 VNs are reduced due to condensing the degree-2 VNHs, while the degree-2 VNCHs remain unchanged. The same can be observed, even more significantly, for degree-2 CNs because the condensed degree-2 CNs that are not connected to any VNCH represent a major percentage of the original CNs. Condensing both VNHs and CNs results in much higher degree (i.e., complex) nodes at a relatively low percentage. One could further refine the pruning by a maximum CN degree, which only allows condensing CNs if the resulting CN degree is below a given limit.

## IV. RESULTS AND COMPLEXITY ANALYSIS

The BER performance of polar codes under different decoders, namely SCL, Arıkan's original BP and our proposed LDPC-like BP over the pruned graph is depicted in Fig. 5. Note that we seek to compare the performance of the different decoding algorithms itself, thus, the SCL does not use an additional Cyclic Redundancy Check (CRC). As can be seen, short-to-intermediate length polar codes ($N = 256, 2048$) show a negligible BER performance loss under LDPC-like BP decoding compared to Arıkan's original BP decoder. One potential reason for a gap is the different scheduling in both decoders. However, we observed that this gap decreases with more BP iterations.

**Algorithm 1** Prune **H**-matrix

**Input:**
    $\mathbf{H}_{orig}$.     ▷ original **H**-matrix
**Output:**
    $\mathbf{H}_{pruned}$.     ▷ pruned **H**-matrix

1: $\mathbf{H}_{pruned} \leftarrow \mathbf{H}_{orig}$
2: $\mathbf{H}_{pruned} \leftarrow$ remove_Frozen_VN($\mathbf{H}_{pruned}$)
3: **while** $true$ **do**
4:     $\mathbf{H}_{pruned} \leftarrow$ prune_degree1_CN($\mathbf{H}_{pruned}$)
5:     $\mathbf{H}_{pruned} \leftarrow$ condense_degree1_VNCH($\mathbf{H}_{pruned}$)
6:     $\mathbf{H}_{pruned} \leftarrow$ prune_degree1_VNH($\mathbf{H}_{pruned}$)
7:     $\mathbf{H}_{pruned} \leftarrow$ condense_degree2_VNH($\mathbf{H}_{pruned}$)
8:     $\mathbf{H}_{pruned} \leftarrow$ condense_degree2_CN($\mathbf{H}_{pruned}$)
9:     **if** size($\mathbf{H}_{pruned}$) does not change **then**
10:         **return** $\mathbf{H}_{pruned}$
11:     **end if**
12: **end while**

Table III: Complexity comparison between Arıkan's original and our proposed LDPC-like BP decoder at a fixed BER = $10^{-4}$ and $N_{it,max} = 200$.

| Code | BP Decoder | $I_{avg}$ | $c_{ac}$ | $m_{cv}$ | $s_{syn}$ |
|---|---|---|---|---|---|
| $\mathcal{P}(256,128)$ | Arıkan | 3.8 | $1.6 \times 10^4$ | $3.9 \times 10^4$ | 60.8 |
| | LDPC-like | 10.9 | $0.4 \times 10^4$ | $2.7 \times 10^4$ | 21.8 |
| $\mathcal{P}(2048,1024)$ | Arıkan | 9.5 | $4.2 \times 10^5$ | $1 \times 10^6$ | 209 |
| | LDPC-like | 31.5 | $1.1 \times 10^5$ | $0.1 \times 10^6$ | 63 |

For complexity comparison, the following criteria are used:
- **Average number of iterations performed** $I_{avg}$: denotes the average number of iterations conducted, given a maximum threshold $N_{it,max}$. For the results shown in Table III, $N_{it,max} = 200$ is used.
- **Average number of active CNs** $c_{ac}$: denotes the average number of CNs activated over the whole BP iterations.
- **Average number of passed messages** $m_{cv}$: denotes the average number of messages interchanged between different nodes over the whole BP iterations.
- **Average latency / number of time steps** $s_{syn}$: denotes the average number of time steps required to decode (i.e., synchronization steps in a parallel implementation and, thus, proportional to the latency).

As depicted in Table III, BP decoding of polar codes over the pruned graphs of the LDPC-like code consumes on average more BP iterations than Arıkan's original BP decoder of the same polar code. However, as Arıkan's original BP factor graph is composed of $\log_2 N$ stages, this, still gives our proposed decoder an advantage (i.e., the pruned graph is composed of *one* stage only). In the original BP decoder, $2 \cdot \log_2 N$ stages need to be *synchronized* per iteration, thus on average less synchronization steps are needed in our proposed decoder. Furthermore, the average number of active CNs throughout the BP iterations is less for our proposed decoder, as a result of the new view of the factor graph and the new decoding schedule which made efficient graph pruning possible. However, one may argue that the CNs in our pruned graphs are of higher degree (i.e., complexity) than the original factor graph. To take that into account, the average number of iterated messages through the graph is compared, and is shown to be significantly less for our proposed decoder as well. Thus from an implementation perspective, our proposed decoder is less demanding in terms of memory requirements, when compared to Arıkan's original BP decoder.

Table II: Dimensions [and density] of the **H**-matrix and the degree profiles before and after pruning.

| Code | dim($\mathbf{H}_{before}$) [den($\mathbf{H}_{before}$)] dim($\mathbf{H}_{after}$) [den($\mathbf{H}_{after}$)] | $\lambda(Z)_{before}$ (VND) $\lambda(Z)_{after}$ (VND) | $\rho(Z)_{before}$ (CND) $\rho(Z)_{after}$ (CND) |
|---|---|---|---|
| $\mathcal{P}(256,128)$ | $2048 \times 2304$ [$\approx 0.11\%$] | $0.075 + 0.4Z^1 + 0.525Z^2$ | $0.4Z^1 + 0.6Z^2$ |
| $\mathcal{P}(256,128)$ | $361 \times 489$ [$\approx 0.71\%$] | $0.103 + 0.205Z^1 + 0.365Z^2 +$ $0.16Z^3 + 0.08Z^4 + 0.013Z^7 +$ $0.029Z^8 + 0.032Z^9 + 0.013Z^{15}$ | $0.103Z^1 + 0.404Z^2 + 0.308Z^3 + 0.08Z^4 +$ $0.019Z^7 + 0.029Z^8 + 0.032Z^9 + 0.026Z^{15}$ |
| $\mathcal{P}(2048,1024)$ | $22528 \times 24576$ [$\approx 0.01\%$] | $0.054 + 0.4Z^1 + 0.546Z^2$ | $0.4Z^1 + 0.6Z^2$ |
| $\mathcal{P}(2048,1024)$ | $3792 \times 4816$ [$\approx 0.075\%$] | $0.075 + 0.149Z^1 + 0.345Z^2 + 0.242Z^3 +$ $0.068Z^4 + 0.045Z^5 + 0.008Z^6 + 0.008Z^7 +$ $0.011Z^8 + 0.006Z^9 + 0.003Z^{10} + 0.004Z^{15} +$ $0.005Z^{31} + 0.005Z^{63} + 0.009Z^{127} + 0.019Z^{255}$ | $0.075Z^1 + 0.399Z^2 + 0.242Z^3 + 0.164Z^4 +$ $0.049Z^5 + 0.01Z^6 + 0.007Z^7 + 0.012Z^{15} +$ $0.005Z^{16} + 0.005Z^{17} + 0.009Z^{31} + 0.005Z^{63} +$ $0.019Z^{127}$ |

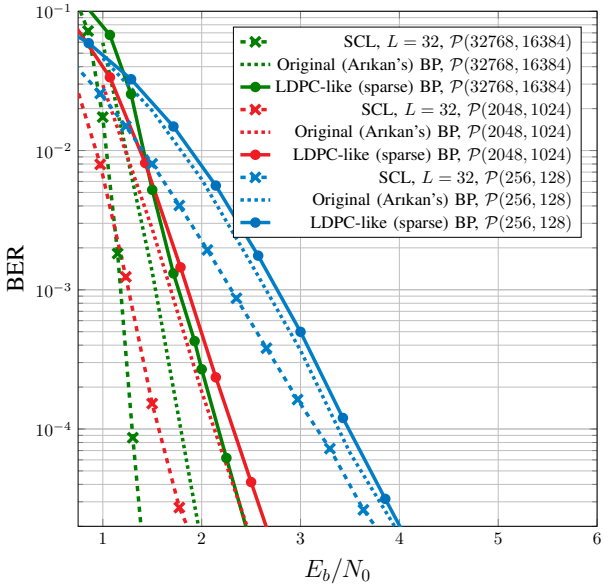

Fig. 5: BER comparison between SCL, Arıkan's original BP and our proposed LDPC-like BP decoder of $\mathcal{P}(256,128)$, $\mathcal{P}(2048,1024)$ and $\mathcal{P}(32768,16384)$ codes; $N_{it,max} = 200$; no CRC used.
Note: As shown in Table III, one full iteration in Arıkan's original BP decoder has higher complexity than one iteration of the LDPC-like decoder. When fixing the decoding complexity, we observed that the BER performance gap between both decoders diminishes.

## V. Conclusion

We show that polar codes can be decoded by a *conventional* LDPC decoder over a sparse graph of a corresponding LDPC-like code. By carefully pruning the resulting **H**-matrix, we achieve a significant reduction in the size of the **H**-matrix, resulting in a faster convergence behavior. The source code for the pruning steps conducted is provided online [11] for testing further possible pruning ideas. Furthermore, we show that there is only a negligible BER performance loss for short-to-intermediate length polar codes, whereas this loss diminishes for more BP iterations. We believe that this approach opens up many new – theoretical and practical – research directions for polar BP decoding, as many tools such as density evolution or EXIT charts may be applied much easier. This may also open door to new and, probably, more efficient hardware implementation, given the latency reduction provided here. In a future step, this could help to identify improved frozen bit positions (i.e., code design) especially tailored to iterative polar decoding. Another possible research area may be to find ways of considering the CRC during iterative decoding.